\documentclass[12pt]{article}
\usepackage{amssymb}
 \usepackage[all,cmtip]{xy}

\def\hybrid{\topmargin 0pt      \oddsidemargin 0pt
        \headheight 0pt \headsep 0pt
        \voffset=-0.5cm
        \hoffset=-0.25in
        \textwidth 6.75in
        \textheight 9.5in       
        \marginparwidth 0.0in
        \parskip 5pt plus 1pt   \jot = 1.5ex}
\catcode`\@=11
\def\marginnote#1{}

\newcount\hour
\newcount\minute
\newtoks\amorpm
\hour=\time\divide\hour by60 \minute=\time{\multiply\hour by60 \global\advance\minute by-\hour}
\edef\standardtime{{\ifnum\hour<12 \global\amorpm={am}%
        \else\global\amorpm={pm}\advance\hour by-12 \fi
        \ifnum\hour=0 \hour=12 \fi
        \number\hour:\ifnum\minute<10 0\fi\number\minute\the\amorpm}}
\edef\militarytime{\number\hour:\ifnum\minute<10 0\fi\number\minute}

\def\draftlabel#1{{\@bsphack\if@filesw {\let\thepage\relax
   \xdef\@gtempa{\write\@auxout{\string
      \newlabel{#1}{{\@currentlabel}{\thepage}}}}}\@gtempa
   \if@nobreak \ifvmode\nobreak\fi\fi\fi\@esphack}
        \gdef\@eqnlabel{#1}}
\def\@eqnlabel{}
\def\@vacuum{}
\def\draftmarginnote#1{\marginpar{\raggedright\scriptsize\tt#1}}
\def\draftlabel#1{{\@bsphack\if@filesw {\let\thepage\relax
   \xdef\@gtempa{\write\@auxout{\string
      \newlabel{#1}{{\@currentlabel}{\thepage}}}}}\@gtempa
   \if@nobreak \ifvmode\nobreak\fi\fi\fi\@esphack}
        \gdef\@eqnlabel{#1}}
\def\@eqnlabel{}
\def\@vacuum{}
\def\draftmarginnote#1{\marginpar{\raggedright\scriptsize\tt#1}}

\def\draft{\oddsidemargin -.5truein
        \def\@oddfoot{\sl preliminary draft \hfil
        \rm\thepage\hfil\sl\today\quad\militarytime}
        \let\@evenfoot\@oddfoot \overfullrule 3pt
        \let\label=\draftlabel
        \let\marginnote=\draftmarginnote
   \def\@eqnnum{(\theequation)\rlap{\kern\marginparsep\tt\@eqnlabel}%
\global\let\@eqnlabel\@vacuum}  }

\def\numberbysection{\@addtoreset{equation}{section}
        \def\theequation{\thesection.\arabic{equation}}}

\def\underline#1{\relax\ifmmode\@@underline#1\else
        $\@@underline{\hbox{#1}}$\relax\fi}

\def\titlepage{\@restonecolfalse\if@twocolumn\@restonecoltrue\onecolumn
     \else \newpage \fi \thispagestyle{empty}\c@page\z@
        \def\thefootnote{\fnsymbol{footnote}} }

\def\endtitlepage{\if@restonecol\twocolumn \else  \fi
        \def\thefootnote{\arabic{footnote}}
        \setcounter{footnote}{0}}  

\numberbysection \hybrid

\newcounter{mo}

\newcommand{\ti}[1]{\tilde{#1}}

\newcommand{\vf}{\varphi}
\newcommand{\al}{\alpha}
\newcommand{\be}{\beta}
\newcommand{\ga}{\gamma}
\newcommand{\om}{\omega}
\newcommand{\vth}{\vartheta}

\newcommand{\bfe}{{\bf{e}}}

\newcommand{\Mat}{\hbox{Mat}(N,\mathbb C)}

\newtheorem{predl}{Proposition}[section]

\newtheorem{rem}{Remark}

\def\beq{\begin{equation}}
\def\eq{\end{equation}}
\def\p{\partial}

\newcommand{\mats}[4]{\left(\begin{array}{cc}{#1}&{#2}\\ {#3}&{#4}
\end{array}\right)}

\def\res{\mathop{\hbox{Res}}\limits}

\begin{document}

\setcounter{page}{1}

\date{}
\date{}
\vspace{50mm}

\begin{flushright}
 ITEP-TH-01/15\\
\end{flushright}
\vspace{0mm}

\begin{center}
\vspace{0mm}
%
{\LARGE{Quantum Baxter-Belavin R-matrices and}}
 \\ \vspace{4mm}
{\LARGE{multidimensional Lax pairs for Painlev\'e VI}}
\\
\vspace{10mm} {\large {A. Levin}\,$^{\flat\,\sharp}$ \ \ \ \ {M.
Olshanetsky}\,$^{\sharp\,\natural}$
 \ \ \ {A. Zotov}\,$^{\diamondsuit\, \sharp\, \natural}$ }\\
 \vspace{8mm}

 \vspace{2mm} $^\flat$ -- {\small{\sf 
 NRU HSE, Department of Mathematics,
 Myasnitskaya str. 20,  Moscow,  101000,  Russia}}\\
 \vspace{2mm} $^\sharp$ -- {\small{\sf 
 ITEP, B. Cheremushkinskaya str. 25,  Moscow, 117218, Russia}}\\
 \vspace{2mm} $^\natural$ -- {\small{\sf MIPT, Inststitutskii per.  9, Dolgoprudny,
 Moscow region, 141700, Russia}}\\
\vspace{2mm} $^\diamondsuit$ -- {\small{\sf Steklov Mathematical
Institute  RAS, Gubkina str. 8, Moscow, 119991,  Russia}}\\
\end{center}

\begin{center}\footnotesize{{\rm E-mails:}{\rm\ \
 alevin@hse.ru,\  olshanet@itep.ru,\  zotov@mi.ras.ru}}\end{center}

 \begin{abstract}
The quantum elliptic $R$-matrices of Baxter-Belavin type satisfy the
associative Yang-Baxter equation in ${\rm Mat}(N,\mathbb C)^{\otimes
3}$. The latter can be considered as noncommutative analogue of the
Fay identity for the scalar Kronecker function. In this paper we
extend the list of $R$-matrix valued analogues of elliptic function
identities. In particular, we propose counterparts of the Fay
identities in ${\rm Mat}(N,\mathbb C)^{\otimes 2}$. As an
application we construct $R$-matrix valued $2N^2\times 2N^2$ Lax
pairs for the Painlev\'e VI equation (in elliptic form) with four
free constants using ${\mathbb Z}_N\times {\mathbb Z}_N$ elliptic
$R$-matrix. More precisely, the four free constants case appears for
an odd $N$ while even $N$'s correspond to a single constant.
 \end{abstract}

\newpage

{\small{

\tableofcontents

}}


\section{Introduction and summary}
\setcounter{equation}{0}

In this paper we continue the study of identities for quantum (and
classical) $R$-matrices, which are similar to the elliptic functions
identities for scalar elliptic functions
\cite{Pol,LOZ9}. More concretely, we prove the Fay identities in
 ${\rm
Mat}(N,\mathbb C)^{\otimes 2}$. It allows us to construct
multidimensional Lax pairs for the Painlev\'e VI equation with the
$R$-matrices as matrix elements.

 We start with the list
of properties and identities for elliptic functions, and then give
their $R$-matrix version.
Most of the properties are known from \cite{Baxter,Belavin},
\cite{RicheyT}, \cite{BazhStrog,Takht}, \cite{Pol} and \cite{LOZ9}.

Consider the following functions:
 \beq\label{a01}
 \begin{array}{c}
  \displaystyle{
 \phi(z,u)=\frac{\vth'(0)\vth(z+u)}{\vth(z)\vth(u)}\,,
  }
 \end{array}
 \eq
  \beq\label{a02}
 \begin{array}{c}
  \displaystyle{
 E_1(z)=\frac{\vth'(z)}{\vth(z)}\,,\ \ \ \ \ E_2(z)=-\p_z E_1(z)=\wp(z)-\frac{1}{3}\frac{\vth'''(0)}{\vth'(0)}\,,
  }
 \end{array}
 \eq
where $\vth(z)$ is the odd Riemann theta-function
 \beq\label{a03}
 \begin{array}{c}
  \displaystyle{
\vth(z)=\vth(z|\tau)=\displaystyle{\sum _{k\in \mathbb Z}} \exp
\left ( \pi \imath \tau (k+\frac{1}{2})^2 +2\pi \imath
(z+\frac{1}{2})(k+\frac{1}{2})\right )
  }
 \end{array}
 \eq
and $\wp(z)$ is the Weierstrass $\wp$-function.

Following \cite{We} the function (\ref{a01}) is referred to as the
Kronecker function, and (\ref{a02}) are called the (first and the
second) Eisenstein functions.

 The Kronecker function can be considered as a section of the
Poincar\'{e} bundle  ${\cal P}$ over $\Sigma_\tau\times\Sigma'_\tau$.
Here $\Sigma_\tau$ is the elliptic curve
 \beq\label{a00}
\Sigma_\tau={\mathbb C}/(\mathbb Z +\tau \mathbb Z)\,,~~\Im m\tau>0\,,
\eq
 $\Sigma'_\tau$ -- is its Jacobian  $~(\Sigma'_\tau\sim\Sigma_\tau)$.
The Poincar\'{e} bundle   ${\cal P}$ is a line bundle over $\Sigma_\tau\times\Sigma'_\tau$

 \beq\label{po}
 \xymatrix{
                &\cal{P}\ar[d]            \\
        &\Sigma_\tau\times    \Sigma'_\tau \ar[ld] \ar[rd]\\
 \Sigma_\tau & &    \Sigma'_\tau         }
 \eq
specialized by (\ref{a04}), (\ref{a09}), (\ref{a72}) and
(\ref{a73}).

The properties of theta-function (\ref{a03}) (including Riemann
identities, see \cite{Mum}) provides the following set of properties
and relations for the functions (\ref{a01})-(\ref{a02}):
 \begin{itemize}
 \item {\em Arguments symmetry:}
 \beq\label{a04}
 \begin{array}{c}
  \displaystyle{
\phi(z,u)=\phi(u,z)\,,
  }
 \end{array}
 ~~z\in\Sigma_\tau\,,~u\in \Sigma'_\tau\,,
 \eq
\item {\em Local expansion:}
 \beq\label{a09}
 \begin{array}{c}
  \displaystyle{
 \phi(z,u)=\frac{1}{z}+E_1(u)+\frac{z}{2}(E_1^2(u)-\wp(u))+
 O(z^2)\,,
  }
 \end{array}
 \eq
  \item {\em Residues:}
 %
 \beq\label{a05}
 \begin{array}{c}
  \displaystyle{
\res\limits_{z=0} \phi(z,u)=\res\limits_{u=0}
\phi(z,u)=\res\limits_{z=0} E_1(z)=1\,,
  }
 \end{array}
 \eq
\item {\em Parity:}
 \beq\label{a08}
 \begin{array}{c}
  \displaystyle{
 \phi(-z,-u)=-\phi(z,u)\,,\ \ \ E_1(-z)=-E_1(z)\,,\ \ \
 E_2(-z)=E_2(z)\,,
  }
 \end{array}
 \eq
 \item {\em (Quasi)periodicity properties:}
 %
 \beq\label{a72}
 \begin{array}{c}
  \displaystyle{
 \phi(z+1,u)=\phi(z,u)\,,\ \ \ E_1(z+1)=E_1(z)\,,\ \ \
 E_2(z+1)=E_2(z)\,,
  }
 \end{array}
 \eq
 \beq\label{a73}
 \begin{array}{c}
  \displaystyle{
 \phi(z+\tau,u)=e^{-2\pi \imath u}\phi(z,u)\,,\ \  E_1(z+\tau)=E_1(z)-2\pi \imath\,,\ \
 E_2(z+\tau)=E_2(z)\,,
  }
 \end{array}
 \eq
\item {\em Heat equation:}
 \beq\label{a091}
 \begin{array}{c}
  \displaystyle{
 2\pi \imath\p_\tau\phi(z,u)=\p_z\p_u\phi(z,u)\,,
  }
 \end{array}
 \eq
 \item {\em Derivatives:}
  \beq\label{a74}
 \begin{array}{c}
  \displaystyle{
 \p_u \phi(z,u)=\phi(z,u)(E_1(z+u)-E_1(u))\,,
  }
 \end{array}
 \eq
 \beq\label{a75}
 \begin{array}{c}
  \displaystyle{
 \p_z \phi(z,u)=\phi(z,u)(E_1(z+u)-E_1(z))\,,
  }
 \end{array}
 \eq
\item {\em Fay (trisecant) identity \cite{Fay}:}
 %
  \beq\label{a10}
  \begin{array}{c}
  \displaystyle{
 \phi(x,u)\phi(y,w)=\phi(x-y,u)\phi(y,u+w)+\phi(y-x,w)\phi(x,u+w)\,,
 }
 \end{array}
 \eq
\item {\em Degenerated Fay identities:}
 \beq\label{a11}
  \begin{array}{c}
  \displaystyle{
 \phi(x,z)\phi(x,w)=\phi(x,z+w)(E_1(x)+E_1(z)+E_1(w) -E_1(x+z+w) )\,,
 }
 \end{array}
 \eq
or
 \beq\label{a12}
  \begin{array}{c}
  \displaystyle{
 \phi(x,z)\phi(y,z)=\phi(x+y,z)(E_1(x)+E_1(y)+E_1(z) -E_1(x+y+z) )\,,
 }
 \end{array}
 \eq

 \beq\label{a13}
  \begin{array}{c}
  \displaystyle{
 \phi(x,z)\phi(x,-z)=E_2(x)-E_2(z)=\wp(x)-\wp(z)\,.
 }
 \end{array}
 \eq
 \item {\em Geometric interpretation:} The Kronecker function $\phi(z,u)$ is a section of the Poincar\'{e} bundle $\mathcal{P}$.
 It is a line bundle over $\Sigma_\tau\times\Sigma_\tau$, defined by the conditions
 (\ref{a04}), (\ref{a09}), (\ref{a72}), (\ref{a73}).

 \item {\em Green function:} The Kronecker function is the Green function for the operator $\bar\partial$ in the space
 of one forms $\mathcal{A}^{(1,0)}(\Sigma_\tau)$ with the boundary conditions
 (\ref{a72}) and (\ref{a73}):
 \beq\label{a127}
  \begin{array}{c}
  \displaystyle{
 \bar\partial\phi(z,u)=\delta^2(z,\bar{z})\,.
 }
 \end{array}
 \eq
 \end{itemize}

{\bf Quantum $R$-matrices.} Consider ${\mathbb Z}_N\times {\mathbb
Z}_N$ (Baxter-Belavin's) elliptic $R$-matrix \cite{Baxter,Belavin}
in the fundamental representation (see also \cite{RicheyT}). It is
defined via the finite-dimensional representation of the Heisenberg
group:
  \beq\label{a14}
 \begin{array}{c}
  \displaystyle{
Q,\Lambda\in \hbox{Mat}(N,\mathbb C):\ \ \
Q_{kl}=\delta_{kl}\exp(\frac{2\pi
 \imath}{N}k)\,,\ \ \ \Lambda_{kl}=\delta_{k-l+1=0\,{\hbox{\tiny{mod}}}
 N}\,,\ \ k,l=1,...,N\,,
 }
 \end{array}
 \eq
  \beq\label{a15}
 \begin{array}{c}
  \displaystyle{
\exp(2\pi
\imath\frac{\ga_1\ga_2}{N})Q^{\ga_1}\Lambda^{\ga_2}=\Lambda^{\ga_2}
Q^{\ga_1}\,,\ \ \ga_1\,,\ga_2\in\mathbb Z\,.
 }
 \end{array}
 \eq
Introduce the sin-algebra basis in $\hbox{Mat}(N,\mathbb C)$:
  \beq\label{a16}
 \begin{array}{c}
  \displaystyle{
T_\ga:=T_{\ga_1\ga_2}=\exp(\pi
\imath\frac{\ga_1\ga_2}{N})Q^{\ga_1}\Lambda^{\ga_2}\,,\ \
\ga_1\,,\ga_2=0,...,N-1\,.
 }
 \end{array}
 \eq
The same definition is used for any $\ga\in{\mathbb Z}^{\times 2}$.
Then
  \beq\label{a17}
 \begin{array}{c}
  \displaystyle{
T_\al T_\be=\kappa_{\al,\be} T_{\al+\be}\,,\ \ \
\kappa_{a,b}=\exp\left(\frac{\pi \imath}{N}(\be_1
\al_2-\be_2\al_1)\right)\,,
 }
 \end{array}
 \eq
where $\al+\be=(\al_1+\be_1,\al_2+\be_2)$. The $R$-matrix is defined
as
 \beq\label{a18}
 \begin{array}{c}
  \displaystyle{
R^\hbar_{12}(u)=\sum\limits_{\al\in\, {\mathbb Z}_N\times {\mathbb
Z}_N}\vf_\al(u,\om_\al+\hbar)\,T_\al\otimes T_{-\al}\in
\hbox{Mat}(N,\mathbb C)^{\otimes 2}\,,
 }
 \end{array}
 \eq
 where\footnote{Here $\p_\tau\om_\al=\al_2/N$.}
 \beq\label{a19}
 \begin{array}{c}
  \displaystyle{
\vf_\al(u,\om_\al+\hbar)=\exp(2\pi\imath
u\p_\tau\om_\al)\phi(u,\om_\al+\hbar)\,,\ \ \
\om_\al=\frac{\al_1+\al_2\tau}{N}\,.
 }
 \end{array}
 \eq
The ${\mathbb Z}_N\times {\mathbb Z}_N$ symmetry means that for
$g=Q,\Lambda$
 \beq\label{a191}
 \begin{array}{c}
  \displaystyle{
(g\otimes g) R_{12}^\hbar(u) (g^{-1}\otimes
g^{-1})=R_{12}^\hbar(u)\,.
 }
 \end{array}
 \eq
For $N=1$ the $R$-matrix (\ref{a18}) is the scalar Kronecker
function $\phi(\hbar,u)$ (\ref{a01}). Notice that (\ref{a18}) is
normalized in such a way that the unitarity condition acquires the
form:
 \beq\label{a20}
 \begin{array}{c}
  \displaystyle{
R_{12}^\hbar(u)R_{21}^\hbar(-u)=N^2\phi(N\hbar,u)\phi(N\hbar,-u)
1\otimes 1=N^2(\wp(N\hbar)-\wp(u)) 1\otimes 1\,.
 }
 \end{array}
 \eq
 The latter can be considered as analogue of (\ref{a13}). Here
 $R_{21}(z)=P_{12}R_{12}(z)P_{12}$, where
 \beq\label{a201}
 \begin{array}{c}
  \displaystyle{
P_{12}=\frac{1}{N}\sum\limits_\al T_\al\otimes
T_{-\al}=\sum\limits_{i,j=1}^N E_{ij}\otimes E_{ji}\,,\ \ \
\left(E_{ij}\right)_{kl}=\delta_{ik}\delta_{jl}
 }
 \end{array}
 \eq
 is the permutation operator.
 We also use notation $R^\hbar_{ab}(z)$ which differs from
 (\ref{a20}) by  $T^a_{\al}\otimes T^b_{-\al}=1\otimes ...  1\otimes T_{\al}\otimes 1... 1\otimes T_{-\al}\otimes 1...
 \otimes1$ instead of $T_\al\otimes T_{-\al}$ (i.e. $T_\al$ and $T_{-\al}$ are in the $a$-th and $b$-th components).
  The number of components in the
 tensor product is an integer $\tilde N$.
 It means that $R_{ab}^\hbar$ is considered as an element of $\hbox{Mat}(N,\mathbb C)^{\otimes \tilde
 N}$, i.e. $N^{\tilde N}\times N^{\tilde N}$ matrix.

 {\bf The properties and identities}
(\ref{a05})-(\ref{a12}) have the following analogues for
$R$-matrices:
 \begin{itemize}
 \item {\em Arguments symmetry:}
 \beq\label{r04}
 \begin{array}{c}
  \displaystyle{
R_{12}^{\hbar}(z)=R_{12}^{\frac{z}{N}}(N\hbar)P_{12}\,,
  }
 \end{array}
 \eq
\item {\em Local expansion} in $\hbar$ is the classical limit:
 \beq\label{r09}
 \begin{array}{c}
  \displaystyle{
 R_{12}^\hbar(z)=\hbar^{-1}\, 1\otimes 1+r_{12}(z)+\hbar\, m_{12}(z)+
 O(\hbar^2)\,,
  }
 \end{array}
 \eq
 where $r_{12}(z)$ is the classical (Belavin-Drinfeld \cite{Belavin})
 $r$-matrix:
 \beq\label{r092}
 \begin{array}{c}
  \displaystyle{
r_{12}(z)=E_1(z)\,1\otimes 1 +\sum\limits_{\al\neq
0}\vf_\al(z)\,T_\al\otimes T_{-\al}
 }
 \end{array}
 \eq
and
 \beq\label{r093}
 \begin{array}{c}
  \displaystyle{
m_{12}(z)=\frac{E_1^2(z)-\wp(z)}{2}\,1\otimes 1
+\sum\limits_{\al\neq 0}\exp(2\pi\imath
z\p_\tau\om_\al)\p_u\phi(z,u)\left.\right|_{u=\om_\al}\,T_\al\otimes
T_{-\al}\,.
 }
 \end{array}
 \eq
 Similarly to (\ref{a09}) we have:
 \beq\label{r094}
 \begin{array}{c}
  \displaystyle{
r_{12}^2(z)-2m_{12}(z)=1\otimes 1\, N^2\wp(z)\,,
 }
 \end{array}
 \eq
 i.e. the quantum $R$-matrix is a matrix analogue of the Kronecker function
 (\ref{a01}) while the classical one is the analogue of the first
 Eisenstein function (\ref{a02}).

 \noindent Expansion with respect to $z$ (near $z=0$) is as follows:
 \beq\label{r095}
 \begin{array}{c}
  \displaystyle{
R_{12}^\hbar(z)=\frac{N P_{12}}{z}+R_{12}^{\hbar,(0)}+O(z)\,,
 }
 \end{array}
 \eq
 where\footnote{$R_{12}^{\hbar,(0)}$ appears as a part of the inverse inertia tensor for relativistic tops \cite{LOZ8}.}
\beq\label{r096}
 \begin{array}{c}
  \displaystyle{
R_{12}^{\hbar,(0)}=\sum\limits_\al T_\al\otimes T_{-\al}\,
(E_1(\hbar+\om_\al)+2\pi\imath \p_\tau\om_\al)\,.
 }
 \end{array}
 \eq
  \item {\em Residues}
 %
 \beq\label{r05}
 \begin{array}{c}
  \displaystyle{
\res\limits_{\hbar=0} R^{\hbar}_{12}(z)=1\otimes 1\,,\ \ \
\res\limits_{z=0}R^{\hbar}_{12}(z)=\res\limits_{z=0} r_{12}(z)=N
P_{12}\,,
  }
 \end{array}
 \eq
\item {\em Parity:}
 \beq\label{r08}
 \begin{array}{c}
  \displaystyle{
 R^{\hbar}_{12}(z)=-R^{-\hbar}_{21}(-z)\,,\ \ \ r_{12}(z)=-r_{21}(-z)\,,\ \ \
 m_{12}(z)=m_{21}(-z)\,.
  }
 \end{array}
 \eq
 The $R$-matrix analogue of $E_2(u)=E_2(-u)$ (\ref{a02}) appears as $F_{12}^0(u)=-\p_u
 r_{12}(u)$ (It is natural because $r_{12}(u)$ is the analogue of $E_1(u)$). The classical $r$-matrix is odd. Hence $F_{12}^0(u)$ is even
 matrix function. The same answer follows from the
 local expansions (\ref{a09}), (\ref{r09}):
 $E_2(u)=-\p_u\phi(z,u)\left.\right|_{z=0}$, then  $-\p_u R_{12}^{\,z}(u)\left.\right|_{z=0}=-\p_u
 r_{12}(u)$.
  \item {\em (Quasi)periodicity properties:}
 %

 \beq\label{r723}
 \begin{array}{c}
  \displaystyle{
 R_{12}^\hbar(z+N\om_\ga)=\exp(-2\pi\imath N\hbar\,\p_\tau\om_\ga)\,( T_\ga^{-1}\otimes 1)R_{12}^\hbar(z)( T_\ga\otimes
 1)\,,
  }
 \end{array}
 \eq
  \beq\label{r724}
 \begin{array}{c}
  \displaystyle{
 R_{12}^{\hbar+\om_\ga}(z)=\exp(-2\pi\imath z \p_\tau\om_\ga)\,( T_\ga^{-1}\otimes 1)R_{12}^\hbar(z)(1\otimes
 T_\ga)\,.
  }
 \end{array}
 \eq
In particular,
 \beq\label{r721}
 \begin{array}{c}
  \displaystyle{
 R_{12}^\hbar(z+1)=(Q^{-1}\otimes 1)R_{12}^\hbar(z)(Q\otimes 1)\,,
  }
 \\ \ \\
  \displaystyle{
 R_{12}^\hbar(z+\tau)=\exp(-2\pi\imath\hbar)\,(\Lambda^{-1}\otimes 1)R_{12}^\hbar(z)(\Lambda\otimes
 1)\,,
  }
 \end{array}
 \eq

  \beq\label{r722}
 \begin{array}{c}
  \displaystyle{
 R_{12}^{\hbar+1}(z)=R_{12}^\hbar(z)\,,
\ \ \
 R_{12}^{\hbar+\tau}(z)=\exp(-2\pi\imath z)\,R_{12}^\hbar(z)\,,
  }
 \end{array}
 \eq

 \beq\label{r73}
 \begin{array}{c}
 \displaystyle{
 r_{12}(z+1)=(Q^{-1}\otimes 1)r_{12}(z)(Q\otimes 1)\,,
 }
 \\ \ \\
  \displaystyle{
 r_{12}(z+\tau)=(\Lambda^{-1}\otimes 1)r_{12}(z)(\Lambda\otimes
 1)-2\pi\imath\,1\otimes 1\,.
  }
 \end{array}
 \eq
Let us also rewrite (\ref{r724}) as follows:
  \beq\label{r725}
 \begin{array}{c}
  \displaystyle{
 R_{ab}^{\hbar+1/N}(z_a-z_b)=Q_a^{-1}R_{ab}^\hbar(z_a-z_b)\, Q_b\,,
  }
 \end{array}
 \eq
  \beq\label{r726}
 \begin{array}{c}
  \displaystyle{
 R_{ab}^{\hbar+\tau/N}(z_a-z_b)=\exp(-2\pi\imath \frac{z_a-z_b}{N})\, \Lambda_a^{-1}R_{ab}^\hbar(z_a-z_b)\,
 \Lambda_b\,.
  }
 \end{array}
 \eq
Recall now the $R$-matrix valued Lax matrix for ${\rm g}_{\ti N}$
Calogero-Moser model \cite{LOZ9}:
  \beq\label{r727}
 \begin{array}{c}
  \displaystyle{
\mathcal L(\hbar)=\sum\limits_{a,b=1}^{\ti N} \ti{\mathrm
E}_{ab}\otimes \mathcal L_{ab}(\hbar)\,,\ \ \ \mathcal
L_{ab}(\hbar)=\delta_{ab}p_a\,1_a\otimes
1_b+\nu(1-\delta_{ab})R_{ab}^\hbar(z_a-z_b)\,.
 }
 \end{array}
 \eq
where $\ti{\mathrm E}_{ab}$ is the standard basis of ${\rm gl}_{\ti
N}$:  $(\ti{\mathrm E}_{ab})_{cd}=\delta_{ac}\delta_{bd}$,
$a,b,c,d=1...\ti N$. Then it follows from (\ref{r725})-(\ref{r726})
that
  \beq\label{r728}
 \begin{array}{c}
  \displaystyle{
\mathcal L(\hbar+1/N)={\bf Q}^{-1} \mathcal L(\hbar)\, {\bf Q}\,,
 }
 \\ \ \\
  \displaystyle{
\mathcal L(\hbar+\tau/N)=\exp(-{\bf Z}/N)\,{\bf \Lambda}^{-1}
\mathcal L(\hbar)\, {\bf \Lambda}\, \exp({\bf Z}/N)\,,
 }
 \end{array}
 \eq
where
  \beq\label{r729}
 \begin{array}{c}
  \displaystyle{
{\bf Q}=\bigoplus\limits_{a=1}^{\ti N}\, Q_a\,,\ \ \ {\bf
\Lambda}=\bigoplus\limits_{a=1}^{\ti N}\, \Lambda_a\,,\ \ \ {\bf
Z}=\bigoplus\limits_{a=1}^{\ti N}\, z_a 1_a
 }
 \end{array}
 \eq
are  block diagonal matrices. The number of blocks is $\ti N\times
\ti N$, the size of a block is $N^{\ti N}\times N^{\ti N}$.

\item {\em Heat equation:}
 \beq\label{r091}
 \begin{array}{c}
  \displaystyle{
 2\pi \imath\p_\tau R_{12}^\hbar(z)=\p_z\p_\hbar R_{12}^\hbar(z)\,.
  }
 \end{array}
 \eq
\item {\em Derivatives}\footnote{The identities for derivatives of $R$-matrix with respect to
 the Planck constant and spectral parameter were found in \cite{BazhStrog}
 and \cite{Takht} respectively. Authors of \cite{BazhStrog,Takht} used different normalization of the
 $R$-matrix.
 }{\em
 :}
 \beq\label{r74}
 \begin{array}{c}
  \displaystyle{
 \p_\hbar R_{12}^\hbar(z)=\frac{1}{2}\left(\, r_{12}(z+N\hbar)\, R_{12}^\hbar(z) + R_{12}^\hbar(z)\, r_{12}(z-N\hbar)\, \right)
 }
 \\ \ \\
   \displaystyle{
 +\frac{N}{2}\Big( E_1(z+N\hbar)-E_1(z-N\hbar)-2 E_1(N\hbar)
 \Big) R_{12}^\hbar(z)\,,
 }
 \end{array}
 \eq
 \beq\label{r75}
 \begin{array}{c}
  \displaystyle{
 \p_z R_{12}^\hbar(z)=\frac{1}{2N}\left(\, r_{12}(z+N\hbar)\, R_{12}^\hbar(z) - R_{12}^\hbar(z)\, r_{12}(z-N\hbar)\, \right)
 }
  \\ \ \\
   \displaystyle{
 +\frac{1}{2}\Big( E_1(z+N\hbar)+E_1(z-N\hbar)-2 E_1(z)
 \Big) R_{12}^\hbar(z)\,.
 }
 \end{array}
 \eq
 %
\item {\em The Fay identity in $\Mat^{\otimes 3}$ \cite{Aguiar,Pol,LOZ9}:}
   \beq\label{r101}
 \begin{array}{c}
  \displaystyle{
 R^\hbar_{ab}
 R^{\hbar'}_{bc}=R^{\hbar'}_{ac}R_{ab}^{\hbar-\hbar'}+R^{\hbar'-\hbar}_{bc}R^\hbar_{ac}\,,\
 \ \ R^\hbar_{ab}=R^\hbar_{ab}(z_a-z_b)\,.
 }
 \end{array}
 \eq
 Both parts of the identity are elements of $\Mat^{\otimes 3}$. It
 was used in \cite{LOZ9} for constructing higher-dimensional Lax
 pairs for Calogero-Moser models. Here we will prove another analogue of
 (\ref{a10}) -- in $\Mat^{\otimes 2}$.
 \item {\em The Fay identity in $\Mat^{\otimes 2}$:}
   \beq\label{r102}
 \begin{array}{l}
  \displaystyle{
 R^\hbar_{12}(z)
 R_{21}^{\hbar'}(-w)=}
 \end{array}
 \eq
 $$
  \begin{array}{c}
  \displaystyle{
 N\phi(N\hbar',\frac{z\!-\!w}{N}+\hbar'\!-\!\hbar)\, R_{12}^{\hbar-\hbar'}(z\!+\!N\hbar')
 -N\phi(N\hbar,\frac{z\!-\!w}{N}+\hbar'\!-\!\hbar)\, R_{12}^{\hbar-\hbar'}(w\!+\!N\hbar)
 }
\\ \ \\
  \displaystyle{
+N\phi(-w,\frac{z\!-\!w}{N}+\hbar'\!-\!\hbar)\,
R_{12}^{\frac{z-w}{N}}(w\!+\!N\hbar)
-N\phi(-z,\frac{z\!-\!w}{N}+\hbar'\!-\!\hbar)\,
R_{12}^{\frac{z-w}{N}}(z\!+\!N\hbar')\,.
 }
  \end{array}
 $$
 The scalar analogue of this identity is obtained as follows: apply
 (\ref{a10}) (with $x=\hbar$,
 $y=\hbar'$) to $\phi(\hbar,z)\phi(\hbar',-w)$, and then apply (\ref{a10}) once again to the obtained
 r.h.s.. Then we get the scalar analogue of r.h.s. of (\ref{r102}).

\item {\em Degenerated Fay identities in $\Mat^{\otimes 3}$
(\ref{r101}):}
 \beq\label{r11}
  \begin{array}{c}
  \displaystyle{
 R^\hbar_{ab}
 R^{\hbar}_{bc}=R^{\hbar}_{ac}r_{ab}+r_{bc}R^\hbar_{ac}-\p_\hbar
 R_{ac}^\hbar\,,
 }
 \end{array}
 \eq
 \beq\label{r120}
  \begin{array}{c}
  \displaystyle{
 R_{ab}^\hbar(z) R_{bc}^{\hbar'} (-z)=R_{ac}^{\hbar',(0)}
 R_{ab}^{\hbar-\hbar'}(z)+R_{bc}^{\hbar'-\hbar}(-z)R_{ac}^{\hbar,(0)}+N
 F_{bc}^{\hbar'-\hbar}(-z) P_{ac}\,,
 }
 \end{array}
 \eq
where $F_{ab}^\hbar(u)=\p_u R_{ab}^\hbar (u)$ and
$R_{ab}^{\hbar,(0)}$ is from (\ref{r095})-(\ref{r096}).

\item {\em Degenerated Fay identities in $\Mat^{\otimes 2}$
(\ref{r102}):}
 \beq\label{r12}
  \begin{array}{c}
  \displaystyle{
 R_{12}^\hbar(z)R_{21}^\hbar(-w)=N\phi(\frac{z-w}{N},N\hbar)\,(\,r_{12}(z+N\hbar)-r_{12}(w+N\hbar)\,)
 }
 \\ \ \\
  \displaystyle{
 +N\phi(\frac{w-z}{N},z)\, R_{12}^{\frac{z-w}{N}}(z+N\hbar)-N\phi(\frac{w-z}{N},w)\, R_{12}^{\frac{z-w}{N}}(w+N\hbar)
 }
  \\ \ \\
  \displaystyle{
 +N^2 1\otimes 1\, \phi(\frac{z-w}{N},N\hbar)\,
 (E_1(N\hbar)-E_1(N\hbar+\frac{z-w}{N}))\,,
 }
 \end{array}
 \eq
and
 \beq\label{r13}
  \begin{array}{c}
  \displaystyle{
 R_{12}^\hbar(z)R_{21}^{\hbar'}(-z)= N\phi(\hbar'-\hbar,-z)\,(\, r_{12}(z+N\hbar)-r_{12}(z+N\hbar') \,)
 }
 \\ \ \\
  \displaystyle{
 -N\phi(\hbar'-\hbar,N\hbar)\,
 R_{12}^{\hbar-\hbar'}(z+N\hbar)+N\phi(\hbar'-\hbar,N\hbar')\,
 R_{12}^{\hbar-\hbar'}(z+N\hbar')
 }
  \\ \ \\
  \displaystyle{
 +N^2 1\otimes 1\, \phi(\hbar'-\hbar,-z)\, ( E_1(z)-E_1(z+\hbar-\hbar')
 )\,.
 }
 \end{array}
 \eq
 \item \emph{Geometric interpretation.}
   Due to the quasi-periodicities (\ref{r723})-(\ref{r722}) the $R$-matrix have
the following geometrical interpretation.
 Let $V_1$ ($V_2$) be a
rank $N$ and degree one vector bundle over elliptic curve $\Sigma^{(1)}_\tau$ with coordinate $z_1$
($\Sigma^{(2)}_\tau$ with coordinate $z_2$). Consider the bundle $V_1\boxtimes V_2$ over
 $\Sigma^{(1)}_\tau\times\Sigma^{(2)}_\tau$.
Let $Aut_{{\rm PGL}(N)}(V_1\boxtimes V_2)$ be the
automorphism group of the bundle (the gauge group). The sections
 $\Gamma(Aut_{{\rm PGL}(N)}(V_1\boxtimes V_2))$ depends only on the anti-diagonal
$\tilde\Sigma_\tau$ of $\Sigma^{(1)}_\tau\times\Sigma^{(2)}_\tau$ with the coordinate $z=z_1-z_2$.
 Let $\tilde\Sigma'_\tau$ be the dual curve,
   $\hbar$ is the coordinate on  $\tilde\Sigma'_\tau$ and ${\cal P}$ is the
Poincar\'{e} bundle ${\cal P}$ over $\tilde\Sigma_\tau\times\tilde\Sigma'_\tau$ (\ref{po}).
Then  the $R$-matrix (\ref{a18}) is a section
$$
R_{12}^\hbar(z)\in\Gamma\left((Aut_{{\rm PGL}(N)}(V_1\boxtimes V_2))\otimes{\cal P}\right)\,.
$$

 \item \emph{Green function.}
Similarly to (\ref{a127}) the $R$-matrix can be considered as the
Green function of $\bar\p$-operator:
 \beq\label{gf1}
 \bar{\p} R_{12}^{\hbar}(z)=NP_{12}\delta^2(z,\bar{z})\,.
 \eq
 \end{itemize}

Properties (\ref{r09})-(\ref{r091}) simply follows from their scalar
counterparts except (\ref{r094}) which follows from the unitarity
condition (\ref{a20}) in the classical limit (\ref{r09}). Identities
for derivatives (\ref{r74}), (\ref{r75}) were obtained in
\cite{BazhStrog,Takht}. Degenerated Fay identities (\ref{r11}),
(\ref{r120}) in $\Mat^{\otimes 3}$ follows from the nondegenerated
one (\ref{r101}) and local expansions (\ref{r09}), (\ref{r095}).

%

{\bf Our main interest} (in this paper) is the Fay identity in
$\Mat^{\otimes 2}$ (\ref{r102}) and its degenerations (\ref{r12}),
(\ref{r13}). We prove them below.
The computational trick is based on the "arguments symmetry"
property (\ref{r04}) and the scalar Fay identities
(\ref{a10})-(\ref{a12}).

\bigskip

{\bf Painlev\'e VI.} As an application of the obtained formulae we
construct higher-dimensional Lax pairs for the Painlev\'e VI
equation. Denote the half-periods of the elliptic curve
$\Sigma_\tau$ as
 \beq\label{a41}
 \begin{array}{c}
  \displaystyle{
 \{\Omega_a\,,\ a=0,1,2,3\}=\{ 0\,,\ \frac{1}{2}\,,\ \frac{1+\tau}{2}\,,\ \frac{\tau}{2}
 \}\,.
  }
 \end{array}
 \eq
The Painlev\'e VI equation in the elliptic form \cite{P6} is
 \beq\label{a42}
 \begin{array}{c}
  \displaystyle{
\frac{d^2u}{d\tau^2}=-\sum\limits_{a=0}^{3}\nu_a^2\wp'(u+\Omega_a)\,.
  }
 \end{array}
 \eq
Let $N$ be an odd (positive) integer. Consider the following pair of
block-matrices\footnote{The coefficient $1/{\sqrt{-2}}$ gives the
normalization of the constants as in (\ref{a42}).}:
 \beq\label{a43}
 \begin{array}{c}
  \displaystyle{
L(\hbar)=\frac12\,\frac{du}{d\tau}\mats{1\otimes 1}{0}{0}{-1\otimes
1}+\sum\limits_{a=0}^3\,\frac{\nu_a}{N\sqrt{-2}}\, \mats{0}{\mathcal
R^{\hbar,a}_{12}(u)}{\mathcal R^{\hbar,a}_{21}(-u)}{0}
  }
 \end{array}
 \eq
  \beq\label{a44}
 \begin{array}{c}
  \displaystyle{
M(\hbar)=\sum\limits_{a=0}^3\,\frac{\nu_a}{N\sqrt{-2}}\,
\mats{0}{\mathcal F^{\hbar,a}_{12}(u)}{\mathcal
F^{\hbar,a}_{21}(-u)}{0}
  }
 \end{array}
 \eq
 where
 \beq\label{a45}
 \begin{array}{c}
  \displaystyle{
\mathcal R^{\hbar,a}_{12}(u)=\exp(2\pi\imath
N\hbar\,\p_\tau\Omega_a)R_{12}^\hbar(u+N\Omega_a)\,,
  }
  \\ \ \\
  \displaystyle{
\mathcal R^{\hbar,a}_{21}(-u)=\exp(-2\pi\imath
N\hbar\,\p_\tau\Omega_a)R_{21}^\hbar(-u-N\Omega_a)\,,
  }
 \end{array}
 \eq
and
 \beq\label{a46}
 \begin{array}{c}
  \displaystyle{
\mathcal F^{\hbar,a}_{12}(u)=\exp(2\pi\imath
N\hbar\,\p_\tau\Omega_a)F_{12}^\hbar(u+N\Omega_a)\,,
  }
  \\ \ \\
  \displaystyle{
\mathcal F^{\hbar,a}_{21}(-u)=\exp(-2\pi\imath
N\hbar\,\p_\tau\Omega_a)F_{21}^\hbar(-u-N\Omega_a)
  }
 \end{array}
 \eq
with
  \beq\label{a47}
 \begin{array}{c}
  \displaystyle{
F_{ab}^\hbar(u)=\p_u R_{ab}^\hbar(u)\,.
  }
 \end{array}
 \eq
The matrices $L(\hbar)\,,M(\hbar)\in\hbox{Mat}(2,\mathbb
C)\otimes\Mat^{\otimes 2}$. Their size equals $2N^2\times 2N^2$.
The Painlev\'e VI equation (\ref{a42}) is equivalent to the
monodromy preserving equation
  \beq\label{a48}
 \begin{array}{c}
  \displaystyle{
\frac{d}{d\tau}
L(\hbar)-\left(\frac{1}{2\pi\imath}\right)\frac{d}{d\hbar}
M(\hbar)=[L(\hbar),M(\hbar)]\,,
  }
 \end{array}
 \eq
where the Planck constant $\hbar$ plays the role of the spectral
parameter (see \cite{LOZ9}).

 For $N=1$ the answer (\ref{a43}), (\ref{a44}) reproduces the elliptic $2\times 2$ Lax
pair proposed in \cite{Z}.

The Lax pair (\ref{a43}), (\ref{a44}) works for even $N$'s as well.
But the Painlev\'e equation in this case has only one free constant:
 \beq\label{a49}
 \begin{array}{c}
  \displaystyle{
\frac{d^2u}{d\tau^2}=-\nu^2\wp'(u)\,,\ \ \
\nu^2=\sum\limits_{a=0}^3\nu_a^2\,.
  }
 \end{array}
 \eq

\section{Kronecker double series and Baxter-Belavin  $R$-matrix}
\setcounter{equation}{0}

Following idea suggested in \cite{Pol} we derive here the
Baxter-Belavin $R$-matrix as generalization of the Kronecker series.

\textbf{$R$-matrix in Jacobi variables.} Represent the elliptic
curve $\Sigma_\tau$ (\ref{a00}) in the Jacobi form
 $$
C_q=\mathbb{C}/q^{\mathbb{Z}}\,,~~q=\bfe(\tau)=\exp\,2\pi\imath\tau\,.
 $$
Consider the product $C_q\times C_q$ with the coordinates
$s=\bfe(u)$, $t=\bfe(z)$. Instead of the Kronecker function
$\phi(z,u)$ we consider the distribution $g(s,t)$ on the space of
the Laurent polynomials $\mathbb{C}[[s^{-1},t^{-1},s,t]]$. For
$|q|<|t|<1$ it can be represented as the series
 \beq\label{deq0}
g(s,t|\,q)=\sum_{n\in\, \mathbb{Z}}\frac{t^n}{q^ns-1}\,.
 \eq
 If
simultaneously  $|q|<|s|<1$ then
 \beq\label{g}
g(s,t|\,q)=-g^+(s,t|\,q)+g^-(s,t|\,q)\,,~~g^+(s,t|\,q)=\sum_{i,n\geq
0}s^iq^{in}t^n\,,~~ g^-(s,t|\,q)=\sum_{i,n< 0}s^iq^{in}t^n
 \eq
  or
 \beq\label{deq2}
g(s,t|\,q)=1-\frac{1}{1-t}-\frac{1}{1-s}+g^-(s,t)-\sum_{i,n>
0}s^iq^{in}t^n\,.
 \eq
 In the domain $|q|<|t|<1$ and $|q|<|s|<1$ we
have
 \beq\label{deq5}
g(s,t|\,q)|_{s=\frac{1}{2\pi\imath}\ln u,\,t=\frac{1}{2\pi\imath}\ln
z}=\phi(z,u)\,.
 \eq
The distribution $g(s,t|\,q)$ has the  properties analogous to
(\ref{a04})-(\ref{a08}). In particular,
 \beq\label{deq1}
g(s,t|\,q)=g(t,s|\,q)\,.
  \eq
It follows from (\ref{g}) that
 \beq\label{deq3}
g(s^{-1},t^{-1}|\,q)=-g(s,t|\,q)+\delta(t)+\delta(s)-2\,,
 \eq
where  $\delta(s)$ is the distribution on the space of the Laurent polynomials\\
 $\mathbb{C}[t,t^{-1}]=\{\psi(t)=\sum_lc_lt^l\}$, defined by the functional
$\langle\delta,\psi\rangle=Res|_{t=0}\psi(t)$ and represented by the formal series
 \beq\label{def}
\delta(t)=\sum_{n\in\mathbb{Z}}t^n\,.
 \eq
The analog of the quasiperiodic property (\ref{a73}) is the
following. The distribution $g(s,t)$ is a solution of the difference
equation on $t$ (the Green function) variable
 \beq\label{deq4}
sg(s,tq|\,q)-g(s,t|\,q)=\delta(t)-1\,.
 \eq
 It defines the continuation of $g(s,t|\,q)$ from the annulus $|q|<|t|<1$ to $\mathbb{C}^*$.
Due to (\ref{deq1}) the similar equation can be written with respect to the $s$ variable.

Let $\eta=\bfe(\hbar)$. The $R$-matrix (\ref{a18}) takes the
following form in variables $(s,t,\eta)$:
 \beq\label{rkr}
R^\hbar_{12}(s)=\sum\limits_{\al\in\, {\mathbb Z}_N\times {\mathbb
Z}_N}s^{\al_2/N}g(s,\om_\al+\hbar)\,T_\al\otimes T_{-\al}=
 \eq
$$
\sum\limits_{\al\in\, {\mathbb Z}_N\times {\mathbb
Z}_N}s^{\al_2/N}\left(\sum_{m,n}\bfe(n\al_1/N)q^{n(m+\al_2/N)}\eta^ns^m
\right)\,T_\al\otimes T_{-\al}\,.
$$
%
It plays the role of the Green function for the difference operator
%
 \beq\label{gf2}
\eta\, (\Lambda\otimes 1)\,R^\hbar_{12}(sq)\,(\Lambda^{-1}\otimes
1)-R^\hbar_{12}(s) =(\delta(s)-1) P_{12}\,.
 \eq
\bigskip
\textbf{Kronecker double series \cite{We}}\\
The distribution $g(s,t|\,q)$ (and  $\phi(z,u)$) can be represented as a
Kronecker double series. Consider the lattice in $\mathbb{C}$
$$
W=\{\gamma=m+n\tau\,,~m,n\in\mathbb{Z}\}\,.
$$
Represent the argument $u$ of $\phi(z,u)$ as $u=u_1+u_2\tau$ $(u_1,u_2$ are real), and
let
$$
\chi_u(\gamma)=\bfe(-mu_2+nu_1)
$$
be a character of the lattice $W$ ($\chi_u(\gamma)\,:\,W\to S^1$),
parameterized by $u\in\Sigma_\tau$.
 The Kronecker double series is defined as:
 \beq\label{krs}
S(z,u|\,\tau)=\sum_{\ga\in W}\frac{\chi_u(\gamma)}{z+ \ga}\,.
 \eq
From the definition we find that
 \beq\label{qpk}
\begin{array}{c}
  \displaystyle{
 S(z+1,u|\,\tau)=\bfe(u_2)S(z,u|\,\tau)\,,
  }
 \\ \ \\
  \displaystyle{
 S(z+\tau,u|\,\tau)=\bfe(-u_1)S(z,u|\,\tau)\,.
  }
 \end{array}
  \eq
It was proved in \cite{We} that $S(z,u|\,\tau)$ is related to the Kronecker function as
 \beq\label{k1}
S(z,u|\,\tau)=\bfe(u_2 z)\phi(z,u)\,,
 \eq
or in the Jacobi coordinates
 \beq\label{kro}
S(t,s|\,q)=t^{u_2}g(s,t|\,q)\,.
 \eq

\bigskip

Let us now pass to the $R$-matrix and describe it in terms of the
 Kronecker double series $S(z,u|\,\tau)$ (\ref{krs}).

 Define the lattice $W$ by the two generators $(\alpha_1/N+\hbar_1, (\alpha_2/N+\hbar_2)\tau)$,
  where $\hbar=\hbar_1+\hbar_2\tau$, $\hbar_{1,2}\in\mathbb R$.
The corresponding character of $W$ is
  \beq\label{nr1}
\begin{array}{c}
  \displaystyle{
\chi_{(m,n)}(\al,\hbar)=
\bfe\left(-m(\alpha_2/N+\hbar_2)+n(\alpha_1/N+\hbar_1)\right)\,.
 }
 \end{array}
  \eq
 Then the $R$-matrix (\ref{a18}) is defined in terms of
the  Kronecker double series (\ref{krs}) as
 \beq\label{nrm}
\begin{array}{c}
  \displaystyle{
{R}^\hbar_{12}(z)=\bfe(-\hbar_2z)\sum_{(m,n)\in\mathbb{Z}\oplus\mathbb{Z}}\frac
{\sum\limits_{\al\in\, {\mathbb Z}_N\times {\mathbb
Z}_N}\chi_{(m,n)}(\al,\hbar)\,T_\al\otimes T_{-\al}}
{z+m+n\tau}\,.
 }
 \end{array}
  \eq
The quasi-periodicities (\ref{r721}), (\ref{r722}) now become evident.
It follows from (\ref{k1}) that the singular behavior $z,\hbar\to 0$  of
 this representation is in agreement with (\ref{r05}).

 We pass from ${R}^\hbar_{12}(z)$ to the modified matrix
 $$
 \tilde{{R}}^\hbar_{12}(z)=\bfe(\hbar_2z){R}^\hbar_{12}(z)\,.
 $$
 It satisfies the Yang-Baxter equation and has the quasi-periodicities
 $$
  \begin{array}{c}
  \displaystyle{
  \tilde{R}_{12}^\hbar(z+1)=\bfe(\hbar_2)(Q^{-1}\otimes 1) \tilde{R}_{12}^\hbar(z)(Q\otimes 1)\,,
  }
 \\ \ \\
  \displaystyle{
  \tilde{R}_{12}^\hbar(z+\tau)=\bfe(\hbar_1)\,(\Lambda^{-1}\otimes 1)
  \tilde{R}_{12}^\hbar(z)(\Lambda\otimes
 1)\,,
  }
 \end{array}
 $$
 (compare with (\ref{r721})). In contrast with (\ref{r722}) $\tilde{R}$ is not holomorphic
 in $\hbar$ and  is double-periodic.
 \begin{rem}
The representation (\ref{nrm}) means that the elliptic $\tilde{R}$-matrix
is represented as the averaging of the Yang matrix $z^{-1}P_{12}$
along the lattice $W$ twisted by the character (\ref{nr1}).
\end{rem}

From (\ref{r09}) we also find the representation for the classical
$r$-matrix:
 $$
\begin{array}{c}
  \displaystyle{
r_{12}(z)=E_1(z)\,1\otimes 1 +\sum_{m,n\in(\mathbb{Z}\oplus\mathbb{Z})\backslash(0,0)}\frac
{\sum\limits_{\al\in\, {\mathbb Z}_N\times {\mathbb
Z}_N}\chi_{(m,n)}(\al,0)\,T_\al\otimes T_{-\al}} {z+m+n\tau}
 }
 \end{array}
 $$
and
 $$
\begin{array}{c}
  \displaystyle{
m_{12}(z)=\frac{E_1^2(z)-\wp(z)}{2}\,1\otimes 1
+\sum_{m,n\in(\mathbb{Z}\oplus\mathbb{Z})\backslash(0,0)}\frac
{\sum\limits_{\al\in\, {\mathbb Z}_N\times {\mathbb
Z}_N}(z+m+n\bar{\tau})\chi_{(m,n)}(\al,0)\,T_\al\otimes T_{-\al}}
{(z+m+n\tau)(\bar{\tau}-\tau)}\,.
 }
 \end{array}
 $$

\section{Derivation of identities}
\setcounter{equation}{0}

 \begin{predl}
The $R$-matrix (\ref{a18}) satisfies the arguments symmetry property
(\ref{r04}).
 \end{predl}
\noindent\underline{\emph{Proof:}} Using definitions (\ref{a201})
and (\ref{a17}) we have
 \beq\label{a501}
 \begin{array}{c}
  \displaystyle{
R_{12}^{\frac{z}{N}}(N\hbar)P_{12}=\frac{1}{N}\sum\limits_{\al,\be}
T_\al T_\be \otimes T_{-\al} T_{-\be}
\,\vf_{\al}(N\hbar,\om_\al+\frac{z}{N}) }
 \\ \ \\
   \displaystyle{
=\frac{1}{N}\sum\limits_{\al,\be} \kappa_{\al,\be}^2\, T_{\al+\be}
\otimes T_{-\al-\be}\, \vf_{\al}(N\hbar,\om_\al+\frac{z}{N})\,.
  }
 \end{array}
 \eq
Since $\kappa_{\al,\be}=\kappa_{\al,\al+\be}$, the property
(\ref{r04}) is equivalent to the following set of $N^2$ identities:
 \beq\label{a502}
 \begin{array}{c}
  \displaystyle{
\frac{1}{N}\sum\limits_{\al} \kappa_{\al,\ga}^2\,
\vf_\al(N\hbar,\om_\al+\frac{z}{N})=\vf_\ga(z,\om_\ga+\hbar)\,,\ \
\forall\,\ga\in{\mathbb Z}^{\times 2}
  }
 \end{array}
 \eq
 or
  \beq\label{a5021}
 \begin{array}{c}
  \displaystyle{
\frac{1}{N}\sum\limits_{\al} \kappa_{\al,\ga}^2\,
\vf_\al(z,\om_\al+\hbar)=\vf_\ga(N\hbar,\om_\ga+\frac{z}{N})\,,\ \
\forall\,\ga\in{\mathbb Z}^{\times 2}\,.
  }
 \end{array}
 \eq
The latter is verified by comparing residues. To do it we also need
the relation for the sums of $N$-th roots of $1$ (it also follows
from $P_{12}^2=1$):
 \beq\label{a503}
 \begin{array}{c}
  \displaystyle{
\sum\limits_{\al} \kappa_{\al,\ga}^2=N^2\delta_{\ga,\,0}\,.
  }
 \end{array}
 \eq
Let us calculate the residue of both parts of (\ref{a502}) at
$\hbar=-\om_\mu$. The answer for the r.h.s. is obviously
$\delta_{\mu,\ga}\exp(2\pi\imath \p_\tau\om_\ga z)$ due to
(\ref{a05}). For the l.h.s. we have:
 \beq\label{a504}
 \begin{array}{c}
  \displaystyle{
\res\limits_{\hbar=-\om_\mu}\frac{1}{N}\sum\limits_{\al}
\kappa_{\al,\ga}^2\,
\vf_\al(N\hbar,\om_\al+\frac{z}{N})=\res\limits_{\hbar=0}\frac{1}{N}\sum\limits_{\al}
\kappa_{\al,\ga}^2\, \vf_\al(N\hbar-N\om_\mu,\om_\al+\frac{z}{N})
  }
  \\ \ \\
    \displaystyle{
\stackrel{(\ref{a72}),(\ref{a73})}{=}\res\limits_{\hbar=0}\frac{1}{N}\sum\limits_{\al}
\kappa_{\al,\ga}^2\, \kappa^2_{\al,-\mu}\exp(2\pi\imath
\p_\tau\om_\mu z)\, \vf_\al(N\hbar,\om_\al+\frac{z}{N})
  }
  \\ \ \\
    \displaystyle{
\stackrel{(\ref{a05})}{=}\frac{1}{N}\sum\limits_{\al}
\kappa_{\al,\ga-\mu}^2\,\exp(2\pi\imath \p_\tau\om_\mu
z)\frac{1}{N}\stackrel{(\ref{a503})}{=}\delta_{\mu,\ga}\exp(2\pi\imath
\p_\tau\om_\mu z)\,.\ \blacksquare
  }
 \end{array}
 \eq

 \begin{predl}
The $R$-matrix (\ref{a18}) satisfies the Fay identity (\ref{r102})
in $\Mat^{\otimes 2}$.
 \end{predl}
\noindent\underline{\emph{Proof:}} Consider
 \beq\label{a510}
 \begin{array}{c}
  \displaystyle{
R_{12}^\hbar(z)R_{21}^{\hbar'}(-w)=-\sum\limits_{\al,\be}
\kappa_{\al,\be}^2 T_{\al+\be}\otimes
T_{-\al-\be}\,\vf_\al(z,\om_\al+\hbar)\,\vf_\be(w,\om_\be-\hbar')=
  }
 \end{array}
 \eq
Here we already used $R_{21}^{\hbar'}(-w)=-R_{12}^{-\hbar'}(w)$.
Apply the Fay identity  (\ref{a10}), then  (\ref{a5021}), and then
(\ref{a10}) again:
 \beq\label{a511}
 \begin{array}{c}
  \displaystyle{
=-\sum\limits_{\al,\be} \kappa_{\al,\be}^2 T_{\al+\be}\otimes
T_{-\al-\be}\,\vf_\al(z-w,\om_\al+\hbar)\,\vf_{\al+\be}(w,\om_{\al+\be}+\hbar-\hbar')
  }
  \\ \ \\
    \displaystyle{
-\sum\limits_{\al,\be} \kappa_{\al,\be}^2 T_{\al+\be}\otimes
T_{-\al-\be}\,\vf_\be(w-z,\om_\be-\hbar')\,\vf_{\al+\be}(z,\om_{\al+\be}+\hbar-\hbar')
  }
 \end{array}
 \eq
 \beq\label{a512}
 \begin{array}{c}
  \displaystyle{
=-N\sum\limits_{\ga}  T_{\ga}\otimes
T_{-\ga}\,\vf_\ga(N\hbar,\om_\ga+\frac{z-w}{N})\,\vf_{\ga}(w,\om_{\ga}+\hbar-\hbar')
  }
  \\ \ \\
    \displaystyle{
+N\sum\limits_{\ga}  T_{\ga}\otimes
T_{-\ga}\,\vf_\ga(N\hbar',\om_\ga+\frac{z-w}{N})\,\vf_{\ga}(z,\om_{\ga}+\hbar-\hbar')
  }
 \end{array}
 \eq
 \beq\label{a513}
 \begin{array}{c}
  \displaystyle{
=N\sum\limits_{\ga}  T_{\ga}\otimes T_{-\ga}\Bigl(\,
-\phi(N\hbar,\frac{z-w}{N}+\hbar'-\hbar)\,\vf_\ga(w+N\hbar,\om_\ga+\hbar-\hbar')
  }
  \\ \ \\
    \displaystyle{
-\phi(w,\hbar-\hbar'-\frac{z-w}{N})\,
\vf_\ga(w+N\hbar,\om_\ga+\frac{z-w}{N})
  }
    \\ \ \\
    \displaystyle{
+\phi(N\hbar',\frac{z-w}{N}+\hbar'-\hbar)\,\vf_\ga(z+N\hbar',\om_\ga+\hbar-\hbar')
  }
    \\ \ \\
    \displaystyle{
+\phi(z,\hbar-\hbar'-\frac{z-w}{N})\,
\vf_\ga(z+N\hbar',\om_\ga+\frac{z-w}{N})\,\Bigr)\,.\ \blacksquare
  }
 \end{array}
 \eq

 \begin{predl}
The $R$-matrices (\ref{a18}) and (\ref{r092}) satisfies the
degenerated Fay identities (\ref{r12}), (\ref{r13}) in
$\Mat^{\otimes 2}$.
 \end{predl}
\noindent\underline{\emph{Proof:}} We begin with (\ref{r12}).
Consider
 \beq\label{a515}
 \begin{array}{c}
  \displaystyle{
R_{12}^\hbar(z)R_{21}^{\hbar}(-w)=-\sum\limits_{\al,\be}
\kappa_{\al,\be}^2 T_{\al+\be}\otimes
T_{-\al-\be}\,\vf_\al(z,\om_\al+\hbar)\,\vf_\be(w,\om_\be-\hbar)\,.
  }
 \end{array}
 \eq
Subdivide it into two parts: $\sum_{\al,\be}=\sum_{\al\neq
-\be}+\sum_{\al= -\be}$. The first part is transformed as in the
previous Proposition (via (\ref{a10}), then  (\ref{a5021}), and then
(\ref{a10}) again)
 \beq\label{a516}
 \begin{array}{c}
  \displaystyle{
\sum_{\al\neq -\be}=-\sum\limits_{\al\neq -\be} \kappa_{\al,\be}^2
T_{\al+\be}\otimes
T_{-\al-\be}\,\vf_\al(z-w,\om_\al+\hbar)\,\vf_{\al+\be}(w,\om_{\al+\be})
  }
  \\ \ \\
    \displaystyle{
-\sum\limits_{\al\neq -\be} \kappa_{\al,\be}^2 T_{\al+\be}\otimes
T_{-\al-\be}\,\vf_\be(w-z,\om_\be-\hbar)\,\vf_{\al+\be}(z,\om_{\al+\be})
  }
 \end{array}
 \eq
 \beq\label{a517}
 \begin{array}{c}
  \displaystyle{
=...=-N\phi(\frac{z-w}{N},N\hbar)\sum\limits_{\ga\neq 0}
T_\ga\otimes T_{-\ga}\, \vf_\ga(w+N\hbar,\om_\ga)
  }
  \\ \ \\
    \displaystyle{
-N\phi(\frac{w-z}{N},w)\sum\limits_{\ga\neq 0} T_\ga\otimes
T_{-\ga}\,\vf_\ga(w+N\hbar,\om_\ga+\frac{z-w}{N})
  }
  \\ \ \\
    \displaystyle{
+N\phi(\frac{z-w}{N},N\hbar)\sum\limits_{\ga\neq 0} T_\ga\otimes
T_{-\ga}\, \vf_\ga(z+N\hbar,\om_\ga)
  }
  \\ \ \\
    \displaystyle{
+N\phi(\frac{w-z}{N},z)\sum\limits_{\ga\neq 0} T_\ga\otimes
T_{-\ga}\,\vf_\ga(z+N\hbar,\om_\ga+\frac{z-w}{N})
  }
 \end{array}
 \eq
By adding (and subtracting) scalar terms ($1\otimes 1$) to each line
one obtains the first and the second lines of (\ref{r12}). The input
to the scalar part should be summed up together with
 \beq\label{a518}
 \begin{array}{c}
  \displaystyle{
\sum_{\al=-\be}=1\otimes 1\sum\limits_\al
\vf_\al(z,\om_\al+\hbar)\,\vf_\al(-w,\om_\al+\hbar)
  }
  \\
  \displaystyle{
\stackrel{(\ref{a12})}{=}1\otimes 1\sum\limits_\al
\vf_\al(z-w,\om_\al+\hbar)\,
(E_1(z)\!-\!E_1(w)+E_1(\hbar+\om_\al)\!-\!E_1(z-w+\hbar+\om_\al))\,.
  }
 \end{array}
 \eq
The latter expression is transformed via (\ref{a5021}) for $\ga=0$
 $$
\sum\limits_{\al}
\vf_\al(z-w,\om_\al+\hbar)=N\phi(N\hbar,\frac{z-w}{N})
 $$
and its derivative (\ref{a74}), (\ref{a75}) with respect to $\hbar$:
 $$
  \begin{array}{c}
  \displaystyle{
\sum\limits_{\al} \vf_\al(z-w,\om_\al+\hbar)\,
(E_1(z-w+\hbar+\om_\al)-E_1(\hbar+\om_\al))
 }
 \\
  \displaystyle{
=N^2\phi(N\hbar,\frac{z-w}{N})\,\left(E_1(N\hbar+\frac{z-w}{N})
-E_1(N\hbar)\right)\,. }
 \end{array}
 $$
This finishes the proof of (\ref{r12}). The identity (\ref{r13}) can
be derived similarly. Equivalently, (\ref{r13}) follows from
(\ref{r12}) by using the properties (\ref{r04}) and (\ref{r08}).
$\blacksquare$

\section{Higher-dimensional elliptic Lax pairs for Painlev\'e VI}
\setcounter{equation}{0}




Different types of matrix-valued Lax pairs for Painlev\'e equations
are known (see e.g. \cite{JM,Dubrovin,LOZ2}). In this section we
construct $R$-matrix valued generalization of the elliptic $2\times
2$ Lax pair suggested in \cite{Z}.

 \begin{predl}
The Painlev\'e VI equation in the elliptic form (\ref{a42}) is
equivalent to the monodromy preserving equation (\ref{a48}) with the
Lax pair (\ref{a43})-(\ref{a47}) and the elliptic $R$-matrix
(\ref{a18}) for odd $N$.
 \end{predl}
\noindent\underline{\emph{Proof}}
 is similar to the one given in \cite{Z} for the scalar
($N=1$) case.
First, notice that $\frac{d}{d\tau}L(\hbar)=\frac{du}{d\tau}\p_u
L(\hbar)+\p_\tau L(\hbar)$, where the last term is the derivative by
explicit dependence on $\tau$. It is canceled out by
$\frac{1}{2\pi\imath}\frac{d}{d\hbar} M(\hbar)$ due to the heat
equation (\ref{r091}) $2\pi\imath\,\p_\tau \mathcal
R^{\hbar,a}_{bc}(u)=\p_\hbar \mathcal F^{\hbar,a}_{bc}(u)$.

Denote
 \beq\label{a701}
 \begin{array}{c}
  \displaystyle{
L^a= \mats{0}{\mathcal R^{\hbar,a}_{12}(u)}{\mathcal
R^{\hbar,a}_{21}(-u)}{0} \,,\ \ \
 M^a= \mats{0}{\mathcal
F^{\hbar,a}_{12}(u)}{\mathcal F^{\hbar,a}_{21}(-u)}{0}
  }
 \end{array}
 \eq
The main statement which we need to verify is that for $a\neq b$
 \beq\label{a702}
 \begin{array}{c}
  \displaystyle{
[L^a,M^b]+[L^b,M^a]=0\,,
  }
 \end{array}
 \eq
i.e. the input to $[L(\hbar), M(\hbar)]$ comes only from
$[L^a,M^a]$. Indeed, it follows from the unitarity condition
(\ref{a20}) that
 \beq\label{a703}
 \begin{array}{c}
  \displaystyle{
\mathcal R^{\hbar,a}_{12}(u) \mathcal
R^{\hbar,a}_{21}(-u)=R_{12}^\hbar(u+N\Omega_a)R_{21}^\hbar(-u-N\Omega_a)=N^2(\wp(N\hbar)-\wp(u+N\Omega_a))\,.
  }
 \end{array}
 \eq
Differentiating (\ref{a703}) with respect to $u$ we get
 \beq\label{a704}
 \begin{array}{c}
  \displaystyle{
\mathcal
F^{\hbar,a}_{12}(u) \mathcal R^{\hbar,a}_{21}(-u)-\mathcal
R^{\hbar,a}_{12}(u) \mathcal
F^{\hbar,a}_{21}(-u)=-N^2\wp'(u+N\Omega_a)\,.
  }
 \end{array}
 \eq
This identity provides the equation of motion. Notice that in order
to have all four constants $N$ should be odd since
$\wp'(u+N\Omega_a)=\wp'(u+\Omega_a)$ in this case. If $N$ is even
then $\wp'(u+N\Omega_a)=\wp'(u)$, and we have only one constant as
in (\ref{a49}).

To prove (\ref{a702}) let us recall that in the scalar case this
followed from
 \beq\label{a705}
 \begin{array}{c}
  \displaystyle{
\vf_a(\hbar,u+\Omega_a)f_b(\hbar,-u-\Omega_b)-f_b(\hbar,u+\Omega_b)\vf_a(\hbar,-u-\Omega_a)
  }
  \\ \ \\
  \displaystyle{
\vf_b(\hbar,u+\Omega_b)f_a(\hbar,-u-\Omega_a)-f_a(\hbar,u+\Omega_a)\vf_b(\hbar,-u-\Omega_b)=0\,,
  }
 \end{array}
 \eq
where $$f_a(z,u+\Omega_a)=\exp(2\pi\imath\p_\tau\Omega_a
\hbar)\p_w\phi(\hbar,w)\left.\right|_{w=u+\Omega_a}$$ is the scalar
analogue of $\mathcal F^{\hbar,a}_{12}(u)$. The identity
(\ref{a705}) appears from (\ref{a11}) and (\ref{a72})-(\ref{a73}) as
follows:
 \beq\label{a706}
 \begin{array}{c}
  \displaystyle{
\vf_a(\hbar,u+\Omega_a)\vf_b(\hbar,-u-\Omega_b)+\vf_b(\hbar,u+\Omega_b)\vf_a(\hbar,-u-\Omega_a)=
  }
  \\ \ \\
  \displaystyle{
\vf_{a+b}
(\hbar,\Omega_a+\Omega_b)\,(2E_1(\hbar)-E_1(\hbar+\Omega_a-\Omega_b)-E_1(\hbar+\Omega_b-\Omega_a))\,.
  }
 \end{array}
 \eq
The r.h.s. of (\ref{a706}) is independent of $u$. The derivative of
(\ref{a706}) with respect to $u$ gives (\ref{a705}).

Similarly to (\ref{a706}) it follows from the degenerated Fay
identity (\ref{r12}) that
 \beq\label{a707}
 \begin{array}{c}
  \displaystyle{
{\mathcal R}^{\hbar,a}_{12}(u){\mathcal
R}^{\hbar,b}_{21}(-u)+{\mathcal R}^{\hbar,b}_{12}(u){\mathcal
R}^{\hbar,a}_{21}(-u)
 }
 \\ \ \\
  \displaystyle{
=N^2\, 1\otimes 1\,\vf_{a+b}(N\hbar,\Omega_a+\Omega_b)\,
(2E_1(N\hbar)-E_1(N\hbar+\Omega_a-\Omega_b)-E_1(N\hbar+\Omega_b-\Omega_a))\,.
  }
 \end{array}
 \eq
It can be verified directly using (\ref{a72})-(\ref{a73}) which can
be re-written as
$$\phi(z,w+\Omega_a)=\exp(-2\pi\imath
z\p_\tau\Omega_a)\phi(z,w-\Omega_a)\,.$$ The r.h.s. of (\ref{a707})
is scalar and independent of $u$. The derivative of (\ref{a707})
with respect to $u$ gives
 \beq\label{a708}
 \begin{array}{c}
  \displaystyle{
{\mathcal F}^{\hbar,a}_{12}(u){\mathcal
R}^{\hbar,b}_{21}(-u)-{\mathcal R}^{\hbar,a}_{12}(u){\mathcal
F}^{\hbar,b}_{21}(-u)+{\mathcal F}^{\hbar,b}_{12}(u){\mathcal
R}^{\hbar,a}_{21}(-u)-{\mathcal R}^{\hbar,b}_{12}(u){\mathcal
F}^{\hbar,a}_{21}(-u)=0\,.
 }
 \end{array}
 \eq
This identity underlies (\ref{a702}). $\blacksquare$

\subsubsection*{Acknowledgments} We would like to thank N. Slavnov
and A. Zabrodin for useful remarks.
%
The work was done at the Steklov Mathematical Institute RAS, Moscow.
It was supported by RSCF grant 14-50-00005.

\renewcommand{\refname}{{\normalsize{References}}}


 \begin{small}

 \end{small}

\end{document}